\documentclass[]{interact}
\usepackage{epstopdf}
\usepackage{subfigure}

\usepackage[backend=biber, style=authoryear-comp, natbib=true]{biblatex}
\DeclareLanguageMapping{english}{english-apa}
\usepackage[hidelinks]{hyperref}
\usepackage[frozencache=true,cachedir=minted-cache]{minted}

\usepackage[table,xcdraw]{xcolor}
\usepackage{pdflscape}

\theoremstyle{plain}

\theoremstyle{definition}

\theoremstyle{remark}

\usepackage{verbatim}
\usepackage[utf8]{inputenc}

\begin{document}

\articletype{REVIEW ARTICLE}

\title{FAIR Geovisualizations: Definitions, Challenges, and the Road Ahead}


\author{
\name{Auriol Degbelo}
\affil{Institute for Geoinformatics, University of Münster}
auriol.degbelo@uni-muenster.de
}

\maketitle

\begin{abstract}
 The availability of open data and of tools to create visualizations on top of these open datasets have led to an ever-growing amount of geovisualizations on the Web. There is thus an increasing need for techniques to make geovisualizations FAIR - Findable, Accessible, Interoperable, and Reusable. This article explores what it would mean for a geovisualization to be FAIR, presents relevant approaches to FAIR geovisualizations and lists open research questions on the road towards FAIR geovisualizations. \textcolor{black}{The discussion is done using three complementary perspectives: the computer, which stores geovisualizations digitally; the analyst, who uses them for sensemaking; and the developer, who creates them}. The framework for FAIR geovisualizations proposed, and the open questions identified are relevant to researchers working on findable, accessible, interoperable, and reusable online visualizations of geographic information. 
\end{abstract}


\section{Introduction}
\label{sec:introduction}
The momentum gained by the open data movement both in government and research has yielded a large amount of open (geographic) datasets available on the Web. Statistics from Open Data Inception (\url{https://opendatainception.io/}) suggest that there are currently more than 2600 open data portals worldwide. If one considers that the average data portal offers thousands of datasets, this means hundreds of thousands of open datasets to be made sense of. Visualization of these datasets has been identified as a key user need in previous work \citep{degbelo2020needs,Graves2013,graves2014a}. Geovisualizations indeed hold the promise of increasing information transparency \citep{marzouki2017relevance,degbelo2018increasing}, facilitating data-literate citizenry \citep{degbelo2016opening}, and catalyzing citizen participation \citep{fechner2014georeferenced, marzouki2017relevance}, ultimately ``making the world a better place" \citep{Kraak2017}.

Parallel to the developments in the open data landscape, commercial and open-source tools/libraries are increasingly available to create visualizations (e.g. ArcGIS Online, Tableau, D3.js, R). It can thus be expected that (geo)visualizations - next to websites and datasets - will become important contributors to the information overload problem in the digital age. In fact, it can be argued that they are already - witness for example the overwhelming amount of online maps created during the COVID-19 pandemic\footnote{See examples of these maps in \citep{Griffin2020}. For a non-exhaustive collection, see \url{https://www.lrg.tum.de/lfk/service/online-maps-on-corona-covid-19/} (accessed: March 12, 2021).}. Finding online maps on a specific topic is becoming increasingly challenging, though solutions are emerging ad-hoc (primarily in the form of online platforms) to provide entry points for them. Example of these platforms include the D3.js Graph Gallery (\url{https://www.d3-graph-gallery.com/}), DataUSA (\url{https://datausa.io/}), Data-smart city solutions (\url{http://bit.ly/2p23CrS}), ArcGIS StoryMaps (\url{https://storymaps.arcgis.com/}), \textcolor{black}{Observable (\url{https://observablehq.com/collection/@observablehq/maps}), and the R Shiny Gallery (\url{https://shiny.rstudio.com/gallery/})}, to name a few\footnote{All links were last accessed on July 08, 2021.}. The time is ripe to develop techniques making these visualizations first-class in information search processes on the Web. 


\textcite{Safarov2017} identified users of open government data to include citizens, businesses, researchers, developers, non-governmental organizations, and journalists. All these groups can potentially benefit from techniques that make not only websites/datasets about a topic, but also available geovisualizations FAIR - Findable, Accessible, Interoperable, and Reusable. A question at this point is: why care about making online geovisualizations FAIR at all?

From the theoretical perspective, the uniqueness of web maps as geospatial resources can be highlighted from two perspectives mentioned in \citep{lai2021comparative}: map as a \textit{tool} and map as a \textit{representation} of geographic space. First, as a \textit{tool}, web maps are useful to explore the spatial dimensions and interrelationships between phenomena and activities located in space. Thus, they are helpful to create and communicate (visual) stories about geographic phenomena. This storytelling feature is neither present in (raw) datasets nor web services on the spot. Second, as a \textit{form of knowledge representation}, they index information by location in a plane as opposed to using sentences as primary units to organize knowledge. Thus, they enable the retrieval of insight hidden in datasets in a more efficient and effective way. Whether approached from the perspective of a tool that enables story construction, or an artifact that stores geographic knowledge in an effective way, (web) maps are sufficiently distinct from raw datasets and web services to deserve efforts aimed to improve their discoverability.

From the practical perspective, a user survey conducted by \textcite{graves2014a} has reported evidence of stakeholders' interests in the \textit{reuse} of existing online visualizations. Finding these visualizations in the first place is thus critical to meet these users' needs. Yet, the findability of online (geo)visualizations using existing search engines is currently limited. Exemplar questions where state-of-the-art engines would fail to return appropriate answers include: show \textit{all} online maps about the latest earthquake in Christchurch (e.g. for a geologist); show \textit{all} online maps about the demographics of Berlin, Germany (e.g. for an entrepreneur prospecting about opening a new business), show \textit{all} online maps about the kingdom of Prussia (e.g. for a historian who wants to take advantage of map collections available worldwide about their object of study). This suggests that more work on FAIR geovisualizations is currently needed. This article aims to provide a start to discussions about FAIRness issues pertaining to online geovisualizations. Four questions are addressed: 
\begin{itemize}
    \item What does it mean for a geovisualization to be FAIR?
    \item What are approaches relevant to FAIR geovisualizations?
    \item What are unique challenges of FAIR geovisualizations?
    \item What are open questions for FAIR geovisualizations research?
\end{itemize}

\textcolor{black}{`Geovisualization' is a term used with a variety of meanings as indicated by \textcite{Coltekin2018}. It can denote (i) the process of creating interactive visualizations to aid visual thinking and knowledge building during data exploration; (ii) the artifact (e.g. plots, maps, combinations of these) used to support that process; (iii) the artifact resulting from that process (e.g. a map supplemented with graphical annotations that record users' observations during data exploration); or (iv) the academic discipline concerned with the study of the process and the artifacts. The focus of this article is on findable, accessible, interoperable and reusable \textit{artifacts}.} 

\color{black}The two key contributions of the paper are a framework and a research agenda for FAIR online geovisualizations. First, while existing discussions on FAIRness mainly take a computer-centric perspective and focus primarily on technical challenges, a novelty in this work is a holistic discussion of FAIRness. That is, FAIRness is also discussed from the perspective of both the consumer and the producer of the resources to be made FAIR. The discussion of the peculiarities of the three perspectives results into a framework that can be used for (systematic) mapping studies of FAIR geovisualization research. A second novelty of the work is the application of FAIR principles to a new domain, namely that of geovisualizations, to unveil a unique set of challenges and opportunities for GIScience research.  

Relevant literature to the questions above is compiled and critically summarized in this article. Since FAIR geovisualizations is an emerging research topic and not a major theme of any established outlet in GIScience, a systematic review would be premature at this stage. Instead, the article draws on exemplar contributions pertinent to the exposition of the ideas. Section \ref{sec:scene} summarizes the FAIR principles, and presents a scenario for FAIR geovisualizations. Section \ref{sec:background} briefly reports on existing visions/research agendas and highlights their peculiarities from a metaphor-theoretical point of view. As this section shall clarify, FAIR geovisualizations rely on a slightly different metaphor from prevalent ones in geovisualization research. Section \ref{sec:definition} considers the three roles relevant in the context of FAIR geovisualizations (the computer, the analyst, the developer), and elaborates on their uniqueness. Section \ref{sec:challenges} discusses unique challenges of realizing FAIR geovisualizations. Section \ref{sec:roadahead} brings forth some open research questions on the road towards FAIR geovisualizations, Section \ref{sec:limitations} discusses limitations, and Section \ref{sec:conclusion} concludes the article.   


\section{Setting the scene}
\label{sec:scene}
\subsection{What FAIR is}
\label{subsec:whatfairis}
The FAIR principles were originally proposed by \textcite{wilkinson2016fair} in the context of Open Science to emphasize the ability of machines to automatically find and use datasets (or digital assets more broadly), in addition to supporting their reuse by individuals. A wide adoption followed their introduction and with that wide adoption, a variety of interpretations of their meaning emerged. Follow-up articles to the original Wilkinson article tried to clarify the original intent. The following quotes are extracted from some of these. 

\begin{itemize}
\item ``FAIR refers to a set of principles, focused on ensuring that research objects are reusable, and actually will be reused, and so become as valuable as is possible" \citep{Mons2017}.

\item ``The Principles are aspirational, in that they do not strictly define how to achieve a state of `FAIRness', but rather they describe a continuum of features, attributes, and behaviors that will move a digital resource closer to that goal" \citep{Wilkinson2018}. 

\item ``[The Principles] describe characteristics and aspirations for systems and services to support the creation of valuable research outputs that could then be rigorously evaluated and extensively reused, with appropriate credit, to the benefit of both creator and user" \citep{Mons2017}.
\end{itemize}

From the foregoing, FAIR are a set of aspirational principles introduced initially to improve the reuse of research objects. While the principles are useful to research objects, their scope need not be limited to these. They could indeed benefit other types of resources on the Web, namely online geovisualizations. These online geovisualizations may be the outcome of a research process, or not (e.g. many online geovisualizations are typically created by data journalists, not researchers). Realizing FAIR geovisualizations is thus about ensuring that online geovisualizations are reusable, will be reused, and so become as valuable as possible. In essence, the FAIR principles point at the following key considerations discussed in \citep{Jacobsen2020}:

\begin{itemize}
\item Findable: data and metadata must be assigned a globally unique and persistent identifier in order to be found and resolved by computers. In addition, digital resources should be well-described so that they can be accurately discovered.

\item Accessible: data and metadata should be retrieved using a standardized communication protocol. An example of standardized access protocol is the Hypertext Transfer Protocol (HTTP). 

\item Interoperable: data and metadata should be described using a formal language for knowledge representation. (It follows from this definition that interoperability discussions in this article are centered on semantic interoperability aspects, not structural or syntactic aspects).

\item Reusable: data and metadata are released with a clear and accessible usage license.
\end{itemize}
\color{black}


\subsection{FAIR Geovisualizations - A scenario}
\label{subsec:scenario}
\textcolor{black}{Since there is no prior discussion on FAIR online geovisualizations and no elaborated vision of FAIR geovisualizations in the literature, a thought experiment\footnote{According to \textcite{yeates2004thought}, a thought experiment is a device with which one performs an intentional, structured process of intellectual deliberation in order to speculate about (i) potential antecedents for a designated consequent or (ii) potential consequents for a designated antecedent.} called backcasting is used to frame the current discussion.} Following \textcite{yeates2004thought}, backcasting ``involves the description of a definite, specific future situation and, then, moving backwards in time, step-by-step, in as many stages as are considered necessary, from the future to the present. The goal of backcasting is to reveal the mechanism through which the specified future could be attained". Backcasting takes a different approach from forecasting. The latter uses current dominant trends to predict what futures are \textit{likely} to happen. Instead, the former examines \textit{desirable} futures, and how they can be attained. While doing so, a backcasting study may question some suppositions inherent in prevailing perceptions, and open some new options \citep[see][]{Dreborg1996}. \textcolor{black}{Backcasting has been often used for policy planning and analysis (see e.g. \cite{Bibri2018} for a recent review, and \cite{Haslauer2016} for a spatially-explicit backcasting analysis for land-use planning)}. 

\begin{figure}
    \centering
    \includegraphics[scale=0.55]{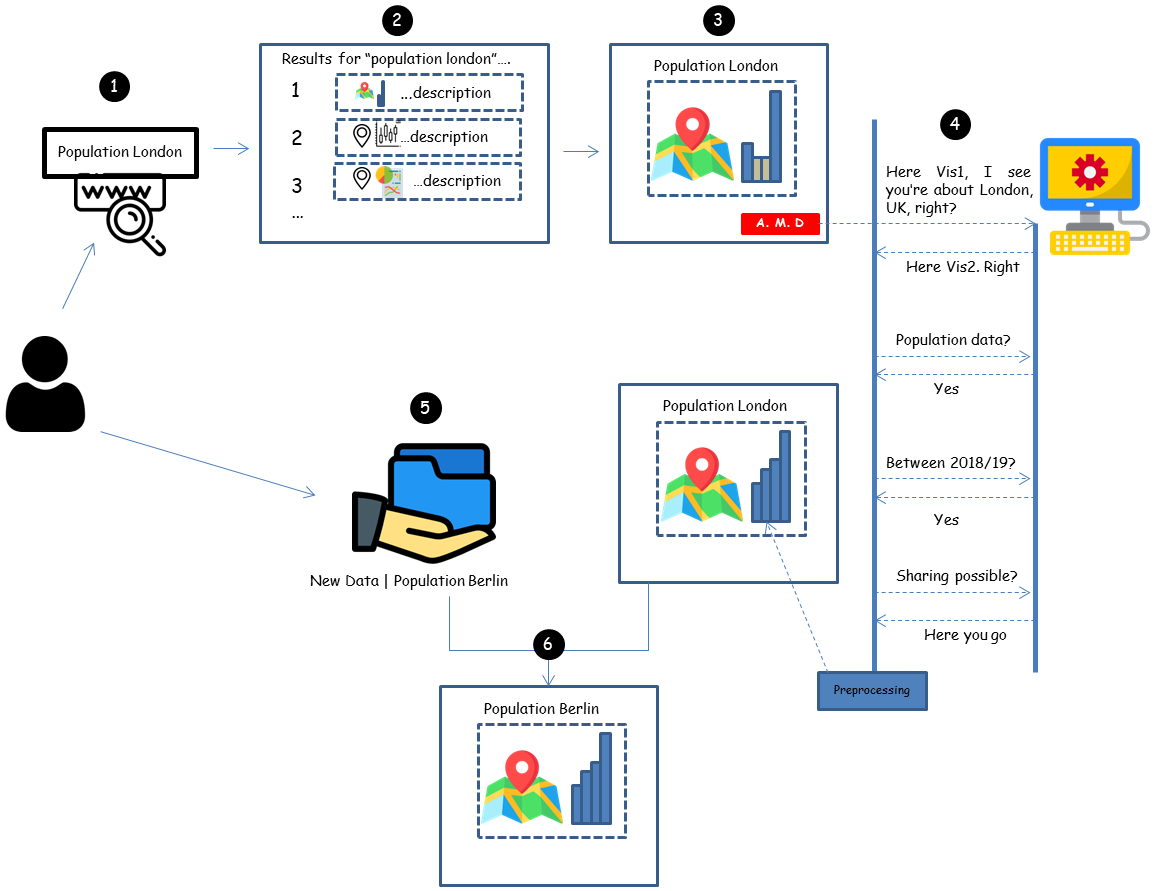}
    \caption{FAIR Geovisualizations at work. 1) the user enters a query in a search engine of their choice; 2) the search engine returns previews and short descriptions of most relevant geovisualizations; 3) the user selects one for closer inspection and realizes (through the grey boxes) that data for the years 2018 and 2019 is missing;  4) the user clicks on the A.M.D. (Add Missing Data) button, and the visualization looks for the missing data on the Web; 5) the user is curious to see if they can visualize data from another geographic area the same way; 6) a simple drag and drop interaction enables the portability of the current visualization template to the new dataset.}
    \label{fig:scenario}
\end{figure}

As an example of \textit{desirable} future\footnote{\color{black}{The scenario below is just one of many possible scenarios and is brought forward to advance the discussion on FAIR geovisualizations. Future work may come up with other equally interesting (if not more interesting) scenarios.}}, consider the following scenario (Figure \ref{fig:scenario}). A data journalist is preparing a new article on population growth in cities, and is at the exploratory stage. They have no background in Cartography and no programming skills. They go through the following steps: 1) they enter a query in a search engine of their choice; 2) the search engine returns previews and short descriptions of most relevant geovisualizations; 3) the data journalist selects one for closer inspection and realizes (through the grey boxes) that data for the years 2018 and 2019 is missing;  4) they click on the A.M.D. (Add Missing Data) button, and the visualization looks for the missing data on the Web; 5) the data journalist is now interested in visualizing data from another geographic area the same way, to compare patterns of the two regions; 6) a simple drag and drop interaction enables the portability of the current visualization template to the new dataset. \textcolor{black}{The focus of this article is on web-based visualizations, of which the generic structure is presented in Listing \ref{list:webgeovis}. For a recent review of web-based visualizations, see \citep{Mwalongo2016}.}

\begin{listing}
\begin{minted}{html}
<!DOCTYPE html>
<html>
<head>
    <title>Geovisualization Title</title>
    <!-- loading relevant libraries, e.g. Leaflet or D3.js -->
    <script src="..."></script> 
</head>

<body>
    Content of the geovisualization......

    <script type="text/javascript">
    // code (e.g.  Leaflet code or D3.js code)
    </script>

</body>

</html>
\end{minted}
\caption{Basic structure of a web-based geovisualization. How to make geovisualizations in this format FAIR?}
\label{list:webgeovis}
\end{listing}
As the reader may have noticed, this scenario necessitates approaching geovisualizations from three distinct and complementary perspectives. In step 2 of the figure, they are \textit{web documents} that are being searched (computer perspective). In steps 3 and 4, they are \textit{software components} that interact with each other (developer perspective). In steps 5 and 6, they are \textit{interfaces} that are manipulated (analyst perspective). Web-based visualizations support the three perspectives (e.g. through the use of \texttt{<meta>} tags for their description and Javascript to add behaviour). \textcolor{black}{From the theoretical point of view, the following observations (drawing on the distinctions introduced by \cite{Larkin1987}) can be helpful to relate the three perspectives.}
\color{black}
\begin{itemize}
    \item At a conceptual level, a geovisualization (as an artifact) is a form of representation of geographic knowledge. \textcite{Larkin1987} distinguished two ways of representing knowledge in general: sentential (i.e. use sentences as a unit to organize information) and diagrammatic (i.e. index information by location in a plane).
    \color{black}
    \item Online geovisualizations are materialized in several, coexistent forms\footnote{The informational and computational equivalence of these three forms are not discussed here.}:
    \begin{itemize}
        \item Sentential, descriptive: this is the case when they are available as declarative statements of facts (e.g. triples, see \cite{Varanka2018}). This form entails considerations from the web documents/computer perspective; 
        \item Sentential, executable: this is the case when they are made available as a software code (e.g. specified using the Vega-Lite grammar, see \cite{Kobben2020}). This form entails considerations from the software/developer perspective;
         \item Diagrammatic: that is, a visual output (e.g. Chart, Map), which may be interactive or not. This form entails considerations from the interface/analyst perspective.
    \end{itemize}
\end{itemize}

The three perspectives (computer, analyst, developer) are discussed in detail in Section \ref{sec:definition}. 

\section{Related work}
\label{sec:background} 
Several research agendas and visions have been suggested touching upon the future of geovisualization research. These research agendas and visions overlap to some extent. A means of highlighting  their subtle differences is through the use of metaphor theory. As \textcite{Lakoff2003} indicated, ``the essence of metaphor is understanding and experiencing one kind of thing in terms of another''.

\subsection{Prevalent metaphors in geovisualization research}
From existing work, at least three metaphors have inspired the progress of geovisualization research. These metaphors, extracted from \citep{maceachren2001research,Kent2000} are: (i) geovisualization as a \textit{knowledge base} (or a database) storing geographical facts; (ii) geovisualization as an \textit{interface} that can support productive information access and knowledge construction activities; and (iii) geovisualization as a \textit{presentation medium} making a specific point and/or persuading the viewer to adopt a particular position. The knowledge base in (i) can take the form of an \textit{image} encoding geographical facts or a \textit{geometric structure} encoding geographic facts (see \cite{Peuquet1988}). 

\color{black}
Existing visions and research agendas have discussed geovisualization as an interface primarily (i.e. metaphor (ii) above). These interfaces can be two-dimensional (as customary with computer screens) or three-dimensional (e.g. a globe or an immersive environment). About 20 years ago, \textcite{maceachren2001research} listed open questions for geovisualization research, starting with the premise that the map is an interface that can support productive information access and knowledge construction activities while retaining its traditional role as a presentation device. \textcite{Roth2013} defined cartographic interaction as the dialog between a human and a map mediated through a computing device. He discussed the what, why, when, who, where, and how of cartographic interaction, and proposed open questions related to these themes. \textcite{Coltekin2017} collected and synthesized community input on what experts considered as persistent challenges in geovisualization. \textcite{Roth2017} discussed methodological aspects of research about user interface aspects (i.e. how to do research about the design and use of interactive maps and visualizations). They also elaborated on research questions in interactive geovisualizations for which more user studies are currently needed.

A few discussions have had a more narrow task focus. \textcite{Griffin2017a} brought forth research opportunities related to the design of geovisualizations as interfaces that are transferable from one context of use to another. The focus of \citep{Andrienko2007} was on how geovisualizations as interfaces can support spatial decision making, while \textcite{Andrienko2010} discussed how they can support visual analysis. \textcite{Robinson2017} looked closely into geovisualization as an interface for the analysis of big spatial data. In \citep{degbelo2018intelligent}, the focus is on making geovisualizations more `intelligent'. The authors brought forth and discussed requirements for intelligent geovisualisations from the perspective of computational user interface design. Finally, \textcite{Thrash2019} focused on geovisualisations for individual pedestrians (they use the term geographic information display) and discuss questions of how these can facilitate navigation through a large-scale, real-world environment. 

\color{black}


\subsection{Another metaphor: geovisualizations as information products}
One core premise of this article is that geovisualizations are \textit{information products} in their own right, in a fashion similar to datasets or websites. \textcolor{black}{From the perspective of metaphor theory, the statement `geovisualizations are information products' is a metaphor, since it involves understanding one domain (i.e. the domain of geovisualizations) in terms of another (i.e. the domain of information products).} Metaphors have entailments, through which they highlight and make coherent some aspects of experience. 


A key entailment of the statement `geovisualizations are information products' is that geovisualizations are \textit{gestalts}. A gestalt is ``an integrated, coherent structure or form, a whole that is different from the sum of the parts'' \citep{Wagemans2012}. That is, \texttt{geoviz = geodata + code + interface}, and the whole is more than the sum of its parts. The information product viewpoint implies more than what the knowledge base metaphor (geodata), the interface metaphor (interface) and the presentation medium metaphor (interface) have covered in isolation. Geovisualizations thus deserve their own treatment in information search. Accordingly, dedicated search mechanisms should be developed to make them findable, just as open data portals (e.g.  Dataverse, Data.gov, European Data portal) or Google Dataset Search \citep{Noy2019} have recently emerged to ease the findability of open datasets.

A second entailment of the sentence `geovisualizations are information products' is that geovisualizations are CONTAINERS (in the metaphor theoretical sense), or in plain language, carriers of informational statements. This entailment is in line with recent proposals to model maps, and more generally visualizations, as a set of statements made by someone at some point in time \citep[see e.g.][]{scheider2014encoding,degbelo2017linked}. 

A third entailment of `geovisualizations are information products' is that geovisualizations are \textit{informational resources} and hence \textit{afford} the FAIR principles from \citep{wilkinson2016fair}. That is, geovisualizations can (or in principle should) be found, accessed, reused, and afford interaction with other information products on the Web. As \textcite{maceachren2004maps} aptly indicated: ``If we accept the premise that maps ... are a useful way of obtaining spatial information, we have the obligation to \textit{facilitate their use as information sources}" (Page 11, emphasis added). The idea that geovisualizations should be made FAIR is easier to grasp when one looks at static images (arguably one of the simplest forms of geovisualization, and one that is already treated as a specific information product by current search engines). 

\textcolor{black}{While geovisualizations afford the FAIR principles, they are yet to be examined in the literature from that perspective. The discussion that follows is an attempt to fill this knowledge gap.}

\section{FAIR Geovisualizations: three perspectives}
\label{sec:definition}
The purpose of this section is to provide a conceptual framework for investigating FAIR geovisualizations. As mentioned above, `geovisualization' can denote an information product, an area of scientific investigation, or a creative process, and the information product viewpoint is of prime interest in this work. 

A geovisualization is an artifact, physical or digital, whose visual properties encode geographic data. \textcite{Roberts2008} proposed that visualizations of geographic information can take one of seven types: maps/cartograms, networks, charts/graphs, tables, symbols, diagrams, and pictures. In other words, geovisualizations subsume maps. Besides, a further distinction needs to be made between machine-readable and non-machine readable geovisualization in the context of information search. An example of the former is a web-based visualization created using D3.js, while an example of the latter is a (PNG or JPEG) image. This implies at least 14 types of geovisualizations to be made findable when \citeauthor{Roberts2008}' seven types are taken into account. Current search engines have dedicated sections for the search of images (e.g. a map in a GIF format), and as such partly address the findability of static or animated geovisualizations. Solutions enabling the findability of interactive web-based geovisualizations about a given topic will be of increasing value in the coming years. 

The application of FAIR principles to geovisualizations requires a clarification of perspectives to remove ambiguities. As \textcite{Roth2013} indicated, cartographic interaction is the dialogue between a map and a human meditated through a computing device. This suggests two key actors affecting a geovisualization during an interaction process: the computer and the human. In addition, the human may be either a producer of the geovisualization or a consumer. In the context of online maps, this distinction matters because the skills required to play these roles are different. A producer requires mastery of a programming language and/or visualization tools (e.g. ArcGIS online, D3.js), while the consumer need not worry about these. \textcolor{black}{There are thus three key roles to consider while investigating FAIR geovisualizations: the computing device (called here `computer'), the consumer (called henceforth `analyst' to borrow the terminology of \citep{Andrienko2010}), and the producer (called henceforth `developer'). The developer may be a prosumer (i.e. produce maps without formal cartographic training), or a trained Cartographer as discussed in \citep{Ipatow2019}. Following the scenario (Figure \ref{fig:scenario}), the analyst may be involved in information analysis tasks and/or information search tasks.} \textcolor{black}{Realizing FAIR geovisualizations implies coping with demands from a computing perspective and addressing needs of analysts. It also puts some constraints on developers of these geovisualizations. While the exact boundaries demarcating these roles are challenging to formally pinpoint, the following informal distinction will be used at this point: the computer stores FAIR geovisualizations digitally, the analyst uses them for sensemaking, and the developer creates them.}

\textcolor{black}{The concurrent presence of these different roles suggests that there are many dimensions of FAIRness, according to the perspective considered. Yet, the literature so far has only taken a computer-centric view of FAIRness issues. To advance our understanding of these and provide a more holistic view, the three perspectives are considered at this point. Interpretations of FAIR according to the role considered, relevant approaches to address the issues arising for each role and scientific communities of interest are considered below. The discussion is done from a knowledge representation perspective (computer); from the perspective of user interface design (analyst); and from the perspective of technical considerations about the publishing of online geovisualizations (developer)}. Table \ref{tab:threeperspectives} summarizes the key takeaways. Since each of the keywords or approaches mentioned in the table would need extensive literature reviews to cover them appropriately, the discussion only points at examples of relevant work and recent reviews, where available.

\begin{landscape}
\begin{table}
\begin{small}
\begin{tabular}{|l|l|p{5.5cm}|p{6cm}|p{7.2cm}|}
\hline
           &                    & \textit{\textbf{Computer's Demands}}                                           & \textit{\textbf{Analyst's Needs}}                                               & \textit{\textbf{Developer's Constraints}}       \\ \hline
\rowcolor[HTML]{FFFFC7} 
\textbf{F} & Definition         & persistent + described                                                            & highlighted                                                             & aliased                           \\ \hline
\rowcolor[HTML]{FFFFC7} 
           & Example approaches & digital object identifiers, containerization; manual- (e.g. tagging, full description), automatic-, and semi-automatic description;  & query auto completion; relevance ranking; visual hierarchy and layout; attention-guiding visualization & query expansion                   \\ \hline
\rowcolor[HTML]{FFFFC7} 
           & Relevant to        & Digital Libraries; Semantic Web                                      & Geographic Information Retrieval                                        & GIScience/Geospatial Semantic Web \\ \hline
\multicolumn{5}{|l|}{}                                                                                                                                                                                               \\ \hline
\rowcolor[HTML]{9AFF99} 
\textbf{A} & Definition         &   compliant (with a standardized communication protocol)                                                        & tailored (to abilities and expertise)                                  & platform- and device-agnostic     \\ \hline
\rowcolor[HTML]{9AFF99} 
           & Example approaches &  https, ftps                                                    & accessible design                                                       & map plasticity                    \\ \hline
\rowcolor[HTML]{9AFF99} 
           & Relevant to        & Web Science                                                          & Human-Computer Interaction                                              & Human-Computer Interaction        \\ \hline
\multicolumn{5}{|l|}{}                                                                                                                                                                                               \\ \hline
\rowcolor[HTML]{CBCEFB} 
\textbf{I} & Definition         & formalized                                                     & compliant (with widespread \textbf{design} conventions)                                     & compliant (with  web mapping standards)        \\ \hline
\rowcolor[HTML]{CBCEFB} 
           & Example approaches & map as statements/knowledge base                                     & symbol stores; map-, text-, and colour brewers                                            & wms, wfs, wcs, sld, html5, css3, svg, rdf/owl, mapML, webGL, x3d/x3dom \\ \hline
\rowcolor[HTML]{CBCEFB} 
           & Relevant to        & GIScience/Cartography \&  GIScience/Geospatial Semantic Web                                   & GIScience/Cartography                                                   & GIScience/Cartography \& GIScience/SDI; Semantic Web                       \\ \hline
\multicolumn{5}{|l|}{}                                                                                                                                                                                               \\ \hline
\rowcolor[HTML]{FFCCC9} 
\textbf{R} & Definition         & licensed (under an open and machine-readable licence)                                         & transformable                                                                & modular + open-source + backward-compatible        \\ \hline
\rowcolor[HTML]{FFCCC9} 
           & Example approaches & JSON hashes, RDF                                                     & template-based design                                                   & declarative frameworks, shelf construction tools, geodocuments, blockchain    \\ \hline
\rowcolor[HTML]{FFCCC9} 
           & Relevant to        & Semantic Web; Open Data                                              & GIScience/Geovisualization \& Information Visualization                                             & GIScience/Geovisualization \& Information Visualization        \\ \hline
\end{tabular}
\end{small}

\caption{A framework for FAIR geovisualizations. The framework features three perspectives: demands from a computing perspective, analyst's needs, and constraints for developers of these geovisualizations. The table also shows examples of approaches to realize these demands/needs/constraints. \textcolor{black}{Finally, the table lists exemplar disciplines and communities that investigate and address the demands/needs/constraints.}}
\label{tab:threeperspectives}
\end{table}
\end{landscape}


\begin{landscape}

\begin{figure}
    \centering
    \includegraphics[scale=0.85]{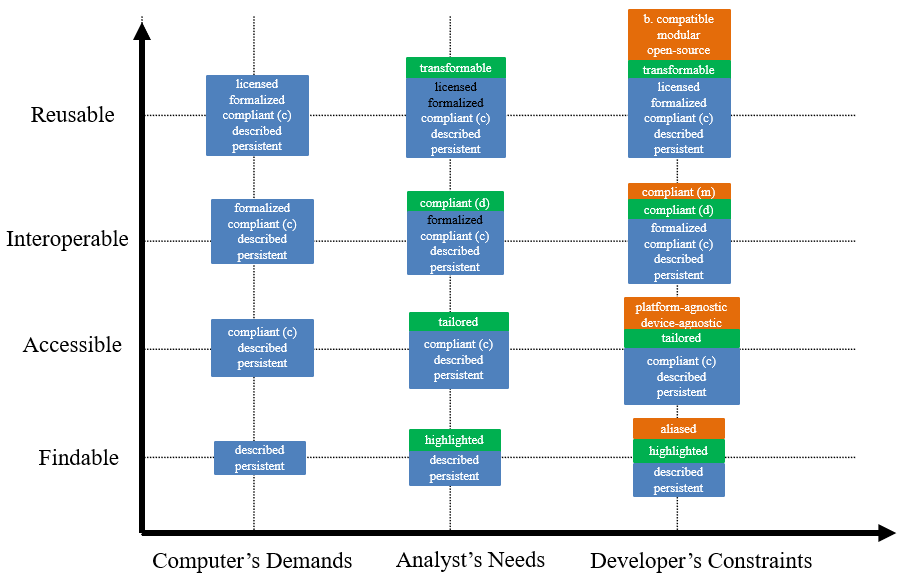}
    \caption{Interdependences between the dimensions of FAIR geovisualizations. The analyst's demands include many of the computer's demands, and the demands of both computer and analyst result in constraints for the developers. Legend: licenced = licenced with a machine-readable open licence; compliant (c) = compliant with standardized communication protocols; compliant (d) = compliant with widespread design conventions; compliant (m) = compliant with web mapping standards. Some aspects are highlighted in black to indicate that there are optional to realize interoperability and reuse from the user's viewpoint.}
    \label{fig:demandsstacked}
\end{figure}

\end{landscape}

\subsection{FAIR Geovisualizations - The computer's demands}
\textbf{F}indable, from the computer's perspective, means `persistent' so that resources can be located over and over again. Web evolution (e.g. API interfaces updates, webpages redesign, server relocation) is the main cause for non-persistence of web urls. Studies on the dynamics of the Web have revealed that death rates for web pages are 20\% after six months \citep{Koehler1999}. Death rates for web sites are 12\% after a six-month period \citep{Koehler1999}, and death rates of Linked Data documents amount to about 5\% for a six-month period \citep{Kafer2013}. Persistence is currently addressed through the use of digital object identifiers (DOIs) to uniquely identify resources, and containerization to encapsulate computational components. Docker is currently the dominant platform for containerization and offers an open-source specification for defining and running multi-container applications (i.e. Docker Compose). Applications based on Docker Compose can be easily deployed on the Web, but the adoption of the technology has been relatively slow, perhaps due to the learning curve required to master it. In addition, since metadata is key to any effective search, \textbf{f}indable from the computer's viewpoint also connotes `described'. Metadata can be generated manually (e.g. place name tags or full-text description), automatically, or semi-automatically. The question of how to produce these metadata has been investigated for scanned maps in the context of digital libraries \citep[e.g.][]{Kiser2018}, and in the context of designing geoportals \citep[e.g.][]{Brovelli2012}. FAIR geovisualizations can learn from, and build upon lessons learned by these and other works. 


\textbf{A}ccessible, from the computer's viewpoint, means `compliant with a standardized communication protocol'. As discussed in \citep{Jacobsen2020}, the purpose of this principle is to provide a predictable way for agents to access resources. Ideally, the protocol should be open, free, and universally implementable. It should allow for authentication and authorization when resources have access restrictions due to ethical, legal or contractual constraints. HTTPs and FTPs are examples of protocols underlying the Web that are compliant with these requirements.

\textbf{I}nteroperable, from the computer's perspective, means encoded in a formal language for knowledge representation (`formalized' for short). A relevant approach here is the idea of map as statements made by somebody somewhere at some time \citep{scheider2014encoding}, also called maps as knowledge base in \citep{Varanka2018}. An early work relying on metadata encoded in the Resource Description Framework (RDF) to improve thematic map search was presented in \citep{Aditya2007}. More recently, \textcite{scheider2014how} illustrated how RDF can be used to describe both the properties of a historical map (e.g. map scale) and its content (i.e. cities, buildings and hills shown on the map). \textcite{degbelo2017linked} showed how the Schema.org vocabulary, which is endorsed by several search engines (e.g. Google, Microsoft, Pinterest, Yandex), can be used to annotate web maps for subsequent search. \textcite{gao2016legendontology} proposed a map legend ontology, and illustrated how the ontology can help answer questions such as ``which map legends contain highway information''? or ``which maps contain both ski areas and camping areas?" Contrary to the works mentioned above, which aimed at describing geovisualizations as \textit{finished information products} in order to improve their search, \textcite{Gould2016}, and \textcite{Huang2019} focused on describing the \textit{process} of producing a geovisualization. The former partly formalized steps of the cartographic generalization process, while the latter formalized the rendering of geometries using ontologies and rules. Finally, MARC \citep{Avram2003,Congress2020} is a format that has been used by projects digitizing paper maps (e.g. \cite{Kiser2018}) in the context of digital libraries. It may be extended, or some of its metadata fields (e.g. projection, type of cartographic material) reused for the description of online web-based geovisualizations.


\textbf{R}eusable, from the computer's viewpoint, means that there is a machine-readable open license available that can be accessed (`licensed' for short). Initiatives to advance work in this area come from the open data community, and include the Open Data Commons' License Service\footnote{\url{https://opendatacommons.org/} (last accessed: March 12, 2021).} and the RDF Creative Commons Licenses\footnote{\url{https://github.com/creativecommons/cc.licenserdf} (last accessed: March 12, 2021).}. The Open Data Commons License Service provides licenses in JSON and an API friendly form, while the RDF Creative Commons Licenses helps describe copyright licenses in RDF.

\subsection{FAIR Geovisualizations - The analyst's needs}
\label{subsec:fairuser}
FAIRness needs from the user's perspective are different from the demands from the computing perspective. For instance, from the analyst's point of view, \textbf{F}indable means `highlighted'. The issue of finding a needle in a haystack is thorny, precisely because the needle resembles all hays around. Highlighting is implemented in various contexts in different ways. To cope with the issue that all queries resemble themselves (syntactically), search engines use an auto-completion feature to highlight queries that are likely to better represent users' intent, and help them avoid spelling mistakes and save time typing. In addition, human-computer interaction and map-making rely on user interface elements arrangement - visual hierarchy and layout \citep[see][]{Tait2018} - to create a ranked order of visual elements so that the most important elements have the greatest visual prominence. For example, a menu item immediately visible on a web page is easier to find that one that needs to be accessed only after browsing. \textcite{Thevenin1999} call the former menu item `first-class' and the latter `second-class'. Finally, relevance ranking is also a means of highlighting items of the information space. It puts some items on top and leaves others at the bottom, signaling to the user where their attention should be directed in priority. \textcite{Hu2015a,Hu2015} used a regression model incorporating nine aspects of matching between input queries and map metadata (e.g. exact or similarity matching on title, description, and place names) to formalize relevance for map search. Their model was applied to the semantic search of maps in ArcGIS online. \textcite{Swienty2008} proposed attention-guiding geovisualization to help users quickly locate and easily decode relevant information through the use of suitable graphical variables. They reported promising results about a computational model to guide users' visual attention to most relevant features during the interaction with a geovisualization.

\textbf{A}ccessible, from the analyst's viewpoint, means `tailored to the user's abilities and expertise'. It is strongly related to \textcite{Shneiderman2000}'s idea of `universal usability': break technology barriers, bridge knowledge gaps, and accommodate user diversity. Example approaches to broaden the accessibility of cartographic products include: the use of multi-layered interfaces to bridge the knowledge gap between novice and experts \citep[see][]{Roth2013}; and the use of multi-modal interaction to engage a multiplicity of user capabilities in the computer-human communication process \citep[see][]{Obrenovic2007}. Challenges of accessible maps were recently discussed in \citep{Froehlich2019}, and recommendations to make web maps more accessible were suggested in \citep{Hennig2017}. From the theoretical point of view, \textcite{Griffin2017a} presented a model for operationalizing map use contexts. The model has four key elements: the \textit{user} (e.g. individual differences, capabilities), the \textit{environment} (e.g. the setting of the map use), the \textit{activity} (e.g. purpose for using the map, actions undertaken while using the map), and the \textit{map} (e.g. representation design, nature of data, display device). Accessible design, as discussed here, primarily addresses the diversity of \textit{user abilities}.


\textbf{I}nteroperable, from the analyst's viewpoint, means that the geovisualization `uses widespread design conventions'. Following \textcite{Roth2011}, a user interacts primarily with four aspects of map features: content, geometries (i.e points, lines, polygons), symbols, and labels. Widespread adoption of conventions is particularly relevant for symbols (map symbology) and labels (map typography). As regards symbology for topographic maps, organizations such as the OrdnanceSurvey\footnote{See \url{https://www.ordnancesurvey.co.uk/mapzone/assets/doc/Explorer-25k-Legend-en.pdf} (last accessed: March 12, 2021).}, the National Park Service\footnote{See \url{https://github.com/nationalparkservice/symbol-library/} (last accessed: March 12, 2021).}, the USGS\footnote{See \url{https://pubs.usgs.gov/gip/TopographicMapSymbols/topomapsymbols.pdf} (last accessed: March 12, 2021).}, and communities such as OpenStreetMap\footnote{See \url{https://wiki.openstreetmap.org/wiki/SymbolsTab} (last accessed: March 12, 2021).} have proposed catalogs of symbols that can be used for application domains such as tourist and leisure information, as well as general-purpose mapping. Reusing symbols from these catalogs can increase interoperability from the user's point of view, since this increases their chances of interacting with `familiar' symbols. \textcolor{black}{Besides, previous work (e.g. \cite{Robinson2013}) has brought forth the concept of a symbol store to promote the sharing of map symbols in a specialized domain (e.g. emergency management)}. Next to topographic maps, symbology for thematic maps needs to specify design conventions for the use of \citeauthor{Bertin1983}'s visual variables during the encoding of geographic data. Approaches to democratize available knowledge regarding such design conventions include (i) recommendations from the literature (e.g. scientific papers, textbooks), and (ii) the use of tools encapsulating (cartography) expertise. As to (i), recommendations for quantitative geodata transformations and effective visualization are available for example in \citep{Kraak2018,Kraak2020}. Guidance for labeling and text placement are available online\footnote{E.g. \url{https://www.arcgis.com/apps/Cascade/index.html?appid=e739c503a1f04d38839834a0fe4ca6d4} (last accessed: Marc 12, 2021).} and in the literature \citep[e.g.][]{Guidero2017}. As to (ii), Colorbrewer \citep{Harrower2003}, the Map Symbol Brewer \citep{schnabel2005map}, and the TypeBrewer\footnote{At the moment of this writing, TypeBrewer is undergoing an update, and is temporarily offline.} are examples of tools suggested to help (web) map makers select/create pertinent color schemes, map symbols, and typographic palettes respectively.  \textcolor{black}{A brewer is a tool that uses mapping principles to offer users a selection of choices for a cartographic representation challenge \citep{Brewer2003}}. ArcGIS and QGIS also provide options to select color/map/typographic palettes during the creation of digital maps. Again, the availability of these tools does not guarantee interoperability per se. Interoperability is truly realized if the symbols provided by all these catalogs are more or less similar, to minimize the element of surprise on the user's side. Collecting empirical evidence on (i) the extent to which the available symbols/palettes are used, and (ii) the extent to which they differ across providers is an open question for research (see Section \ref{subsubsec:interop}).  

\textbf{R}eusable, from the analyst's viewpoint, means `transformable' (or controllable). With respect to the theoretical model of map use contexts from \textcite{Griffin2017a}, reuse from the user's point of view - as discussed here - primarily addresses the diversity of \textit{user activities}. A relevant idea from \textcite{Griffin2017a} is the notion of design transferability, that is, the exploitation of one or more aspects of an existing design for a new application or map use situation. A way of realizing design transferability from a user's perspective is through the use of \textit{templates} (a.k.a. meaningful defaults). These templates are usually available as a feature of a toolkit, and several toolkits have been proposed in the past, for example, Tableau Public\footnote{Tableau Public is sometimes classified as a shelf-configuration user interface (UI) rather than a template editor UI \citep[see e.g.][]{grammel2013survey,Satyanarayan2020}. Here, Tableau Public is mentioned as an example of template-based design software in reference to the ``Show Me" feature that suggests meaningful charts during the visualization creation process.}, IBM ManyEyes \citep{Viegas2007} launched in 2007 and closed in 2015, the GaV Toolkit \citep{VanHo2012}, the Geoviz toolkit \citep{Hardisty2011}, the Geospatial Visual Analytics Toolkit for movement data \citep{Andrienko2008a}, and AdaptiveMaps \citep{degbelo2020datascale} for the visualization of open geographic data.


\subsection{FAIR Geovisualizations - The developer's constraints}
The demands from the computing and user perspective bring about constraints for a developer of FAIR geovisualizations. For example, a constraint imposed by \textbf{F}indable visualizations on the developer is that they should be `aliased' (i.e. items are being referred to in more than one way). \textcite{furnas1987vocabulary} found that the probability of two people spontaneously picking the same term to refer to an entity is less than 20\%, and called this issue the `vocabulary problem'. Recent studies have found equally low rates. \textcite{degbelo2016designing} found that the probability of two providers naming categories for open data search in a similar way is less than 30 \% for European open data catalogs; \textcite{Feitelson2020} reported a probability of less than 7\% regarding developers picking similar variable and function names in their study. As a cure to the vocabulary problem, \textcite{furnas1987vocabulary} suggested that developers use an unlimited number of aliases. Query expansion is the main technique used to mitigate the vocabulary problem. It can be done using (i) terminological associations learned from users' past queries, or (ii) a knowledge base. The former approach has been called `adaptive indices' (see e.g. \cite{Furnas1985}), and the latter typically uses a geographic ontology (e.g. \cite{cardoso2007query,mai2020semantically}), a domain ontology (e.g. \cite{Jiang2018}), or both (e.g. \cite{gaihua2005ontology,Hu2015a}).

A constraint imposed by \textbf{A}ccessible geovisualizations on developers is that they should be `platform- and device-agnostic'. Users can now interact with geovisualizations on different devices (e.g. desktop computers, smartphones, and large-scale displays), using different interaction modalities (touch, eye-movements, sound, freehand gestures, speech). The challenge for the developer is to provide geovisualizations that preserve functionalities and user experience across all these. The idea of interface plasticity \citep{Thevenin1999} and its adaptation to the context of geographic information map plasticity \citep{kray2019map} were proposed to mitigate this issue. Plasticity, as discussed here, primarily addresses the diversity of \textit{user environments} \citep{Griffin2017a}. Designing for plasticity, also called `model-based design' \citep{Meixner2014}, advocates a separation of concerns, so that an interactive system can withstand variations due to the physical characteristics of devices and properties of the software environments. An \textit{abstract model} describes the content of a geovisualization on an abstract level, independent of how it is ultimately rendered to the user. This abstract model is then translated into \textit{one or many concrete geovisualizations}, which respectively fit the constraints imposed by the user's platform (e.g. Android vs iOS) and interaction device (e.g. a tablet or mobile phone). In the context of user interface design, the abstract model has been typically encoded in a machine-readable format, for instance, XML (see \cite{Lacoche2019}), JSON (see \cite{Badam2019}) or the UsiXML (USer Interface eXtensible Markup Language)\footnote{See \url{http://www.usixml.org/} (last accessed: March 12, 2021).}. \textcite{Badam2019} presented BusinessVis, a plastic visualization of the Yelp academic dataset. BusinessVis is web-based and has three views connected through brush-and-link interaction: a map view of businesses, a category treemap, and a rating view showing a list of companies and their user ratings; it supports automated rendering of a JSON-based abstract geovisualization model on both a large display and tablets/smartphones.

A constraint imposed by \textbf{I}nteroperable geovisualizations on developers is that they should be `compliant with (open) web mapping standards'. (Standards here includes both de facto and de jure standards.) In the context of web-based geovisualizations, two types of standards are particularly relevant, namely those from the OGC (Open Geospatial Consortium), and those from the W3C (World Wide Web Consortium). Examples of OGC standards relevant to geovisualizations include the Web Map Service ([WMS] for retrieval of map images), the Web Feature Service ([WFS] for retrieval of geographical features), the Web Coverage Service ([WCS] for retrieval of raster data) and the Styled Layer Descriptor ([SLD] for symbolization and coloring of feature and coverage data). Examples of web standards include HTML5 (for web content structuring and presentation), CSS3 (for styling), and the Scalable Vector Graphics (for resolution-independent graphics). Work is ongoing to bridge the two worlds. At the institutional level, the OGC and the W3C have partnered between 2015 and 2017 to develop a joint note on best practices for the publishing of spatial data on the Web\footnote{See the announcement about the collaboration at \url{https://www.w3.org/2015/01/spatial.html.en}, and the note at \url{https://www.w3.org/TR/sdw-bp/} (both last accessed: March 12, 2021).}. At the conceptual level, \textcite{janowicz2010semantic} proposed a semantic enablement layer for spatial data infrastructures to support the discovery of geospatial content. At the technical level, different adapters were proposed to bridge between the two worlds. These adapters typically support the translation of WFS requests into SPARQL queries (e.g. \cite{VilchesBlazquez2019}), the reverse translation of SPARQL queries into WFS requests (e.g. \cite{zhao2016accessing}), or both (e.g. \cite{jones2014making}). Next to the bridging between different open standards, interoperability between different geovisualization environments is also an important matter of concern. To this end, \textcite{Seo2020} proposed an information model to enable interoperability between geovisualizations in geographic, virtual reality (VR), and augmented reality environments (AR). The information model extends HTML with custom element tags to support the rendering of 2D/3D and VR/AR geo-applications in the browser. Another line of work is developing the Map Markup Language (MapML), a text format for encoding map information for the World Wide Web\footnote{See \url{https://maps4html.org/} (last accessed: March 12, 2021).}. The key objective of MapML is to ``identify the Web map processing that is currently performed by JavaScript libraries which should instead be defined as elements and attributes supported by CSS''\footnote{\url{https://github.com/Maps4HTML/MapML-Proposal} (last accessed: March 12, 2021).}.


A constraint imposed by \textbf{R}eusable geovisualizations on developers is that they should be `modular and open-source'. Modularity is well covered by several web mapping libraries (e.g. R, Leaflet, D3.js). A key idea to facilitate reuse from the developer's perspective is that of \textit{declarative frameworks}. Declarative programming languages specify what the results of a computation should be instead of how the results should be computed. They target programmers for what \textcite{Bostock2013} called `code as Cartography'. Besides, some visualization authoring tools provide developers less conversant in programming with decent expressiveness during the visualization creation process. The Information Visualization literature \citep{grammel2013survey,Satyanarayan2020} distinguishes between \textit{shelf construction tools} (which enable authors to map data fields to visual variables) and \textit{visual builders} (where authors make use of fine-grained user interface elements - marks, glyphs, coordinate systems, and layouts - for more expressiveness during the visualization creation process). Examples of declarative languages for (geo)visualization design include Protovis \citep{Bostock2009}, its successors D3 \citep{Bostock2011} and Vega-lite \citep{Satyanarayan2017}. Examples of shelf construction tools include Tableau Public (\cite[which evolved from Polaris:][]{Stolte2002}) and Voyager \citep{Wongsuphasawat2016}. Examples of visual builders include research prototypes such as Lyra \citep{Satyanarayan2014a} and iVizDesigner \citep{Ren2014}. Furthermore, another important aspect of reuse from the developer's viewpoint is licensing, and at least two useful ideas can be found in existing work: the use of geo-documents, and the use of blockchains. \textcite{doellner2005geospatial} proposed to create geovisualizations as geo-documents that embed digital rights about their components. A geo-document viewer is then in charge of rendering those components of a geovisualization that a user should see (e.g. interaction or animation elements), and hide those they are not allowed to see. \textcite{zhang2020decentralized} proposed a decentralized approach to digital rights management for spatial datasets, and implemented the approach using the open-source blockchain Ethereum. Key features of their approach include: support for digital rights registration, querying, as well as transactions where a software agent asks and explicitly obtains permission before using a spatial dataset; support for web technologies; and support for both raster and vector datasets. Finally, since web mapping libraries used to build online geovisualizations often undergo changes, backward-compatibility of components is an important aspect of reuse from the developers' viewpoint.

\subsection{Summary: what does it mean for a geovisualization to be FAIR?} 
This section has tried to specify what FAIR means for geovisualizations and reported briefly on existing approaches relevant to making geovisualizations FAIR. Two lessons can be learned from the previous sections and Table \ref{tab:threeperspectives}. First, `FAIR' in the context of geovisualization can mean different things depending on the perspective adopted. The twelve senses proposed in Table \ref{tab:threeperspectives} can thus be used by researchers to disambiguate and clarify their standpoint. Second, FAIR geovisualizations provide great potential for interdisciplinary research as their contributions can be relevant to several disciplines and communities. 

\color{black}
Though the aspects of FAIRness have been discussed separately to highlight their unique demands, they are interdependent. Figure \ref{fig:demandsstacked} captures the interdependences. The figure illustrates two key points. First, the concepts of FAIRness involved are multidimensional and the number of dimensions is perspective-dependent as said just above. Second, many of the analyst's demands include the computer's demands to capture the fact that FAIRness from the user's point of view necessitates a solid technological base. \textcolor{black}{The computer's demands and the analyst's needs bring about constraints for the developer.}

\begin{figure}
    \centering
    \frame{
    \includegraphics[scale = 0.8]{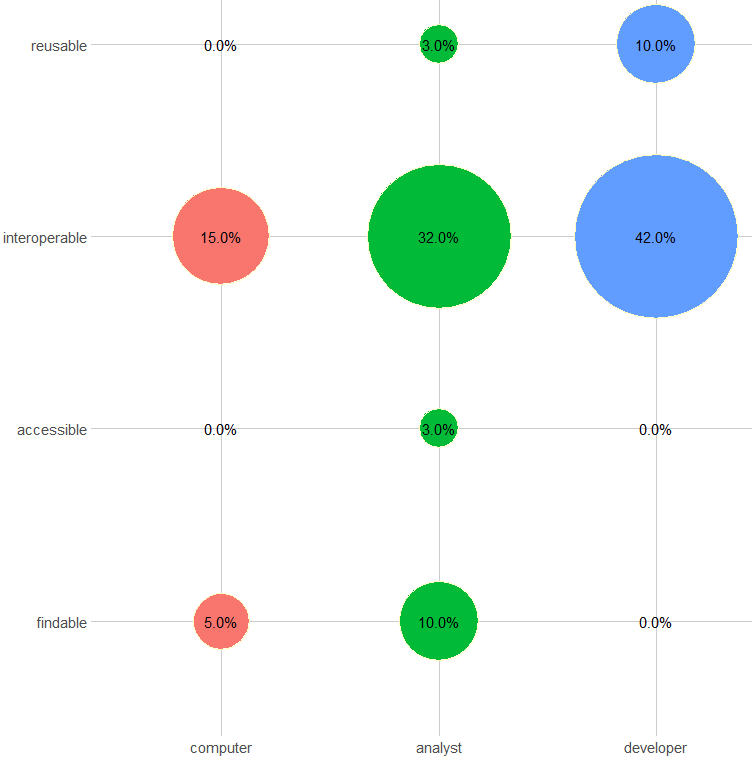}
    }
    \caption{Coverage of research on FAIR geovisualization as of January 2021.}
    \label{fig:systematicmapping}
\end{figure}

As to the usefulness of the framework from Table \ref{tab:threeperspectives}, it can be used to systematically map research on FAIR geovisualizations as it evolves. To illustrate the idea, articles on FAIR geovisualization were downloaded using Taylor \& Francis Online\footnote{\url{https://www.tandfonline.com/}} and the Wiley Online Library\footnote{\url{https://onlinelibrary.wiley.com/}}. These two online catalogs were used because they index the majority of GIScience journals (see the list of journals in the supplementary material). 202 articles were returned from a search done in January 2021 using the keywords: `findable map OR findable geovisualization'; `accessible map or accessible geovisualization'; `interoperable map or interoperable geovisualization'; and `reusable map or reusable geovisualization'. After removing duplicates, review/editorial/encyclopedia articles, 40 articles were found relevant to the issues of knowledge representation, user interface design, or technical considerations of publishing geovisualizations discussed above. The distribution of the topics of these articles is shown in Figure \ref{fig:systematicmapping}. As the figure suggests, the coverage of current research is uneven at the moment. Research on interoperability is well-represented in the sample, and reflects the fact that GIScience work has focused on ontology/knowledge base development, development of empirically-derived guidelines for map design, and development of visualizations based on existing standards (e.g. WMS and WFS). Perhaps unsurprisingly, research on accessibility from the computer perspective is not represented in the sample, as this does not fall within the core of GIScience work. However, research proposing solutions on machine-readable licenses, advancing aliasing to improve semantic search, or highlighting platform/device-independence of geovisualizations has been relatively scarce so far, and could perhaps receive more attention. Including articles from conference proceedings may alter the picture painted here. Thus, these comments on the current coverage of FAIR geovisualization research are tentative. They serve an illustrative purpose for the value of the framework from Table \ref{tab:threeperspectives} only. The data for the figure is available as a supplementary material (Section \ref{sec:supplement}). 

\color{black}

\section{Unique challenges}
\label{sec:challenges}
Having specified how FAIR principles can be applied to geovisualizations, I now turn to the challenges of realizing FAIR geovisualizations. As we shall see, FAIRness applied to web-based geovisualizations presents unique issues for GIScience research. As discussed in Section \ref{sec:definition}, a geovisualization can come in several formats, for example, raster or web documents. Geovisualizations in raster format typically appear in the context of scanned (historical) maps and some challenges they pose have been discussed extensively in previous work \citep[e.g.][]{Chiang2014,Chiang2020}. \textcolor{black}{The focus of this section is on web-based visualizations.} Web visualizations sometimes contain SVG elements, and \textcite{battle2018beagle} have shown that their types (e.g. bar chart, pie chart, map) can be identified with reasonable accuracy (86\%). The challenges of making online web-based geovisualizations FAIR are summarized in Table \ref{tab:uniquechallenges}.

\begin{table}
\begin{small}
\begin{tabular}{|l|p{4.5cm}|p{4.5cm}|p{4.5cm}|}
\hline
           & \textit{\textbf{Computer}}                                                                                              & \textit{\textbf{Analyst}}                                                                                                                                 & \textit{\textbf{Developer}}                                                                                                   \\ \hline
\rowcolor[HTML]{FFFFC7} 
\textbf{F} & persistence is not standalone, but dependent on multiple components (data and libraries); provenance description involves both aspects of data and aspects of processing (e.g. projection, mapping libraries); geovisualization descriptions need to comply with requirements that come with legislation; geovisualizations are composed of many data views & relevance should integrate  unique human factors (e.g. perceptual effectiveness, readability); geovisualizations present multiple perspectives on the data; geovisualizations have multiple usages on the Web  & aliasing, applied to geovisualization content, has no exact limits                                                          \\ \hline
\multicolumn{4}{|l|}{}                                                                                                                                                                                                                                                                                                                                                                                                           \\ \hline
\rowcolor[HTML]{9AFF99} 
\textbf{A} & none    & users' abilities and expertise   are unknown a priori; ability and expertise are multifaced concepts                                                                                                     & device properties (e.g. size,   input modalities) are limiting factors to consistent cross-device   functionality development \\ \hline
\multicolumn{4}{|l|}{}                                                                                                                                                                                                                                                                                                                                                                                                           \\ \hline
\rowcolor[HTML]{CBCEFB} 
\textbf{I} & content modelling as statements has no exact limits, because a geovisualization can communicate many messages at a time; geovisualizations inherently use spatial scale as an organizing dimension for content  & empirically-derived guidelines   are lacking; not all guidelines are amenable to brewers                                                                  & there is a large number of standards to master; protocols to establish mutual understanding during content exchange are still lacking                                                                              \\ \hline
\multicolumn{4}{|l|}{}                                                                                                                                                                                                                                                                                                                                                                                             \\ \hline
\rowcolor[HTML]{FFCCC9} 
\textbf{R} & geovisualization licenses are   inherited, and licensing policies of datasets/libraries may differ                      & the design space of   geovisualizations is complex (e.g. it involves data attributes, data   dimensions, and visual variables)                            & defining/anticipating reuse units is not straightforward; empirical evidence about reuse needs of geovisualization developers is lacking                                                                  \\ \hline
\end{tabular}
\end{small}
\caption{Unique challenges of FAIR geovisualizations from the perspective of knowledge representation (computer), user interface design (analyst) and technical considerations about the publishing of online geovisualizations (developer).}
\label{tab:uniquechallenges}
\end{table}

\subsection{The computer's demands}
As discussed in the previous section, findability necessitates persistence. Given that web-based geovisualizations are composite products, their persistence is dependent on the persistence of their components, notably datasets and libraries. It follows that a persistence strategy for geovisualizations needs an explicit account of these dependencies. In addition, findability of geovisualizations necessitates their description. Description touches primarily on provenance. Aspects of provenance for web pages may include their author, creation date, currency, contact information, and topic. Likewise, aspects of provenance for datasets could include author, creation date, spatial and temporal coverage. A unique feature of geovisualization is that describing them includes data \& web page provenance aspects, but needs to go beyond them to include \textit{aspects of processing} (e.g. map projection used, and web mapping libraries involved in the development of the geovisualization). \textcolor{black}{Furthermore, geovisualizations, qua online products, need descriptions compliant with requirements that come with legislation, e.g. the Americans with Disabilities Act\footnote{\url{https://www.ada.gov/pcatoolkit/chap5toolkit.htm} (last accessed: July 23, 2021).}, or the EU's Web Accessibility Directive (which in turn references the Web Content Accessibility Guidelines)\footnote{\url{https://eur-lex.europa.eu/eli/dir/2016/2102/oj} (last accessed: July 23, 2021).}.} \textcolor{black}{At last, geovisualizations package several views on the data. These views take several forms discussed in \citep{degbelo2020geoinsight}: map, data table, bar chart, pie chart, histogram, Q-Q plots, stacked bars, adjacency diagrams, tree-maps. Rich, fine-granular descriptions need to account for all of these.}

Interoperability requires the use of machine-readable formats. Following \textcite{scheider2014encoding} who distinguished between context (e.g. map scale, spatial extent) and content-based aspects of a geovisualization (i.e. statements that users can extract from the visualization by looking at it), machine-readability can touch on both aspects. Simple examples of describing context and content aspects were presented in \citep{scheider2014encoding}. Yet, the key challenge, when it comes to representing content formally lies in the fact that a geovisualization may provide multiple perspectives (e.g. a bar chart view or a map view) on a given dataset. To this multiplicity of perspectives should be added the different types of meanings conveyed by geovisualizations. As \textcite{maceachren2004maps} pointed out, the meaning of a map can be approached from two angles: denotative and connotative. Denotative meaning is either precisely specified in a map legend or assumed to be part of the normal reader's general map schema; connotative meaning is implicitly conveyed by the map (e.g. connotation of veracity and connotation of integrity). It remains unclear at the moment how these different types of meanings can be represented and linked formally in a geographic knowledge base, for the purpose of (semantic) search. \textcolor{black}{At last, as discussed in \citep{degbelo2020geoinsight}, statements about the content of geovisualizations are granularity-dependent (e.g. some facts pertain only to a given zoom level on a map). This dependence of statements on spatial granularity and the inherent use of spatial scale as an organizing dimension is a distinguishing feature of geovisualizations from other online web documents. Capturing this granularity-dependence in knowledge graphs also presents a unique challenge for semantic interoperability research.}

As to reuse, geovisualization licenses are inherited from those of their components. Since licensing policies of datasets/libraries may differ, a general statement about a geovisualization's reuse policy becomes challenging. Tools to automatically reason about copyright policies of the sub-components and inform about the copyright of the final geovisualization will become increasingly valuable.

\subsection{The analyst's needs}
The primitive spatial query\footnote{\textcolor{black}{There are many more dimensions of spatial querying not discussed here, e.g. objective (e.g. identify, compare), query formulation format (unstructured, formal, visual) and type of user interface feedback (lookup, filtering), see \citep{Andrienko2003}.}} in the context of geographic information retrieval has the form `what, relationship, where' \citep{gaihua2005ontology,cardoso2007query} or `$<$theme$><$relationship$><$location$>$' \citep{Purves2018}. In the context of geodata search, the primitive query takes the form `dataset about, what, where, when' or `dataset about $<$space$><$time$><$theme$>$' \citep{degbelo2020api}. The key difference between geographic information retrieval and geodata search is that the former takes unstructured text as input, while the latter takes a structured description of data items (e.g. in a JSON format) as input. Since web-based geovisualizations are based on HTML that is structured, finding them shares some similarities with dataset search. The primitive query here takes the form `geovisualization about, what, where, when', or `geovisualization about $<$space$><$time$><$theme$>$'. Nonetheless, modeling the notion of \textit{relevance} in this context poses fundamentally new challenges. First and foremost, relevance modeling should integrate perceptual effectiveness, i.e. the ability of a person to retrieve data presented in a chart by decoding a visual representation. Graphical perception studies were conducted to assess, among other things, the effectiveness of bar charts and pie charts \citep{Cleveland1984}, treemaps and wrapped bars \citep{Mylavarapu2019}, line charts and scatterplots \citep{Saket2019}, small multiples for time series \citep{Javed2010}, the differences between interactive tables and interactive geovisualizations \citep{degbelo2018comparison}, and alternative layouts for online map design \citep{Coltekin2009}. These have revealed a strong task-dependency effectiveness for geovisualizations, and general guidelines for visualizations' perceptive effectiveness have not yet emerged. Next to perceptual effectiveness, visual map complexity and map readability are important to the interaction with geovisualizations. The literature has come up with preliminary measures for visual map complexity \citep{Schnur2018} and map readability \citep{harrie2015analytical}. How these could factor in relevance remains an open question. At last, a relevance ranking scheme for online geovisualizations needs to deal with the fact that they have multiple usages. For instance, online geovisualizations can also be used as spatial dialogue platforms (e.g. `Dialog Map', see \cite{fechner2014georeferenced}), platforms for spatial collaboration and data collection (e.g. Ethermap, see \cite{Fechner2015}), tools to persuade (see \cite{Muehlenhaus2013}), or media to display and refine search results. Retrieving online geovisualizations (e.g. thematic maps) that communicate facts about topics is the main purpose of steps 1-2 of the scenario (Figure \ref{fig:scenario}). Finding means to model and distinguish purposes of online geovisualizations is also a unique research challenge.    

Accessibility implies adaptation to the user's abilities and expertise. The key challenge here is that ``for maps presented on the Web, the audience is very much unknown" \citep{Kobben2020}. \textcite{Froehlich2019} listed two high-level challenges for accessible maps: (i) putting accessibility information in existing maps, and (ii) making existing maps more accessible. As to (i), the major impediment is the lack of data about accessibility in general. As to (ii), a key challenge is to model the user's abilities. And the impediment is that `ability' is a multifaceted concept. For instance, \textcite{Froehlich2019} distinguish between sensory abilities (e.g. vision, hearing), physical abilities (e.g. dexterity, mobility), and cognitive abilities (e.g. memory, concentration, language). `Expertise' too is a multifaceted concept. There may be at least four aspects to consider: (i) definition of user expertise, (ii) topics of user expertise, (iii) dimensions of user expertise, and (iv) proxy measures for user expertise. Following \textcite{Downs2015}, an expert can be defined as someone who possesses a certain amount of principled knowledge about a topic. In the case of geovisualization, knowledge is necessary about at least two \textit{topics}: geographic space as well as the operative principles of digital systems. \textit{Dimensions of expertise} may include strategies for problem-solving, the ability to recognize mistakes, and the ability to discriminate between surface (i.e. irrelevant) features and deep (i.e. important) features pertaining to a topic. Finally, \textit{proxy measures} may include experience, performance on specialized tasks, and users' self-assessments. In general, GIScience still lacks means of formally documenting user expertise across evaluation studies of maps and geovisualizations. 

Interoperability implies the use of widespread design conventions, that is, the use of design conventions that can be easily recognized by a broad range of users. There are three key challenges here. First, there is evidence that users do not recognize all map symbols to the same extent, but the causes for these discrepancies are not straightforward to pin down. For instance, \textcite{horbinski2020graphic} conducted a study where they investigated the extent to which users associate map symbols used by map providers (e.g. Google Maps, Open Street Maps, Bing Maps, ArcGIS Maps) to their intended function (e.g. search, geolocation, change of layers). Their results suggested that (i) no one provider achieved high recognition rates on all functions tested, and (ii) recognition rates do not necessarily correlate with the frequency of usage of the services. Second, and as mentioned in \citep{Roth2013,Kray2017}, there is still a lack of (empirically-derived) guidelines specific to cartographic interfaces. The third key challenge is that not all design conventions can be encapsulated into tools (à la ColorBrewer) to facilitate their democratization. Examples here include the use (or not use) of north arrows, recommendations about contrast use and symbol grouping to effectively communicate visual hierarchies \citep{Tait2018}, or effects of the use of visual variables during map icon design \citep{Bell2020}.

Reuse implies giving users the possibility to transform the visualizations as mentioned above, and necessitates some level of adaptivity on the side of the geovisualization. While adaptivity is a truly desirable feature, realizing it in practice proves challenging because the design space of geovisualizations is complex. It involves at least: the  display medium (e.g. mobile phone, large display, desktop computer, tablet computer), interaction modalities (e.g. gaze, speech, touch, mid-air gestures, keyboard, mouse, pen), data attributes (which may be few or many depending on the use case), the measurement scales of the data attributes (e.g. Steven's four levels of measurements), visual variables (up to 12, see \cite{Roth2017a}), and tasks (using \textcite{Brehmer2013}'s typology, the \textit{lower} bound for these is 242). 

\subsection{The developer's constraints}
Findability necessitates the use of aliases to mitigate the vocabulary problem. Previous work reported that `more is not always better for query expansion' \citep{degbelo2019spatial}. Thus, the key challenge here is to find the optimal aliasing strategy, i.e. the one that will maximize user relevance ratings. In the context of data search, aliasing is often performed on the description of the item at hand (e.g. description metadata field about a given dataset). However, in the case of geovisualizations, content is more than mere topical description. Content also covers the statements that a user can extract from a geovisualization by looking at it \citep{scheider2014encoding}. Developing a proper aliasing strategy for geovisualization search is thus an uncharted research area. 

Realizing accessibility from the developer's viewpoint necessitates platform- and device independence (in addition to coping with the diversity of users). Strong constraints are imposed by the device properties (e.g. size, input modalities). In addition, and from the perspective of programming, there is still a need for languages that support plasticity of GI interfaces, so that developers can `build once, deploy anywhere'. These languages should support both declarative (e.g. a map has two layers), and imperative statements (e.g. re-scale map canvas if device size decreases/increases).  

In line with \textcite{Bishr1998}, two geovisualizations \textit{Geoviz1} and \textit{Geoviz2} are said to be interoperable if Geoviz1 can send requests for services \textit{R} to Geoviz2, which then returns responses \textit{S} to Geoviz1, based on a mutual understanding \textit{M}. The exchange of messages is thus key to realize interoperable geovisualizations. Unique challenges here include a large number of standards for developers to master, and the specification of communication protocols that automatically assess `mutual understanding' of content to the exchange. As \textcite{Brodaric2018} indicated, two types of messages are important in the context of interoperability: control messages informing about how the entities interoperate (e.g. error messages, data requests) and content messages containing actual subject matter. HTTP provides a good starting point for the exchange of \textit{control} messages on the Web (e.g. through the use of HTTP headers and HTTP content negotiation). \textcite{Knuth2016} suggested a vision (and preliminary implementation) of using web technologies to exchange rich \textit{content} messages for web media. Their approach relies on the use of HTTP 303 See Other (for redirection to semantic descriptions of media), and the use of RDF as a content description language. Nonetheless, there is still a lack of infrastructure where software agents can autonomously establish the existence of a `mutual understanding' (e.g. our concepts of `River' overlap or are equivalent).

Code reuse is an active research topic in the field of software engineering (see e.g. \cite{Krueger1992,Capilla2019}), and can happen opportunistically (i.e. copy-pasting snippets from existing programs into new programs) or systematically, with the latter more challenging to realize in practice (see \cite{Schmidt1999}). As discussed in \citep{Mikkonen2019}, the emergence of Internet-based developer forums (e.g. Stack Overflow) and open-source software repositories (e.g. GitHub) has catalyzed opportunistic reuse. Nonetheless, the key challenge here is to define/anticipate reuse units, in order to encapsulate them as modules. As \textcite{Brereton1998} indicated: ``Web page developers rarely want to reuse existing pages completely, but they may want to reuse existing page components". This likely holds for geovisualizations. Thus, understanding why developers of geovisualizations engage in reuse enterprises, what they expect (i.e. at which granularity reuse happens more often), and issues they face, is a challenge of geovisualization research. This understanding will help design best practices (relevant for instance to the GIScience teaching curriculum) about `designing for reuse'. 

\subsection{Summary: what are unique challenges of FAIR geovisualizations?}
This section has looked closely into the unique challenges of realizing FAIR geovisualizations. As the section indicated, realizing FAIR geovisualizations demands more of current research regarding provenance description, relevance modelling, user modelling, model-driven development, semantic interoperability, persistence management, licensing policies management and software reuse. \textcolor{black}{As mentioned in Section \ref{subsec:whatfairis}, FAIRness is not an absolute state, but a continuum of behaviours (see also, \cite{DeMirandaAzevedo2020}). That is, ``absolute FAIRness" is not achievable, but online geovisualizations can be made more FAIR, gradually and continually. Also, the challenges listed in this section do not exhibit the same benefit-cost ratio at the time of writing. Addressing the findability of online geovisualizations (steps 1 and 2 of the scenario) seems to offer low-hanging fruits at the moment: the number of online geovisualizations is increasing rapidly and building tools to produce metadata about these so that web-crawlers can semi-automate their retrieval is a task within reach\footnote{See \citep{lai2021comparative} for a recent example of work reporting promising results on semantic metadata generation for online maps.}. If in addition, the metadata is formalized and include machine-readable license information, the computer's demands (Table \ref{tab:threeperspectives}) can already be satisfied to some extent. Accessibility, Interoperability and Reuse challenges from the analyst and developer's viewpoint would need more sustained effort for substantial progress to become visible.}

\section{The road ahead}
\label{sec:roadahead}
Some of the features mentioned in the scenario above (Figure \ref{fig:scenario}) are partly addressed by existing tools. In particular, \textcite{Hu2015a} and \textcite{mai2020semantically}'s work on semantic search for ArcGIS online can provide valuable insights as one moves from search of geovisualizations in one geoportal to their search on the Web as a whole. Yet, they are many open questions raised by the scenario. They are summarized in Table \ref{tab:openquestions}. 

\begin{table}
\begin{small}
\begin{tabular}{|l|p{11cm}|l|l|l|}
\hline
           & \textit{\textbf{Questions}}                                                                                        & \textit{\textbf{C}} & \textit{\textbf{A}} & \textit{\textbf{D}} \\ \hline
\rowcolor[HTML]{FFFFC7} 
\textbf{F} &                                                                                                                    &                     &                     &                     \\ \hline
\rowcolor[HTML]{FFFFC7} 
           & How to generate large-scale semantic annotations for web-based geovisualizations?                                  & x                   & x                   &                    \\ \hline
\rowcolor[HTML]{FFFFC7} 
           & How to enable fine-grained annotation of geovisualization   components?                                            &                    &      x              &                    \\ \hline
\rowcolor[HTML]{FFFFC7} 
           & How to record and share geographic insights across interaction sessions?                                                       & x                   & x                   &                    \\ \hline
\rowcolor[HTML]{FFFFC7} 
           & How to model relevance and incorporate relevance feedback into geovisualization search?                            &                     & x                   & x                   \\ \hline
\rowcolor[HTML]{FFFFC7} 
           & How to realize effective aliasing for geovisualization search?                                                     &                     &                     & x                   \\ \hline
\rowcolor[HTML]{FFFFC7} 
           & How to design mobile search for geovisualizations?                                                                 &                    & x                   & x                   \\ \hline
\rowcolor[HTML]{FFFFC7} 
           & What are developer needs and issues for container-based   geovisualization development?                            &                     &                     & x                   \\ \hline
           
\multicolumn{5}{|l|}{}                                                                                                                                                                            \\ \hline
\rowcolor[HTML]{9AFF99} 
\textbf{A} &                                                                                                                    &                     &                     &                     \\ \hline
\rowcolor[HTML]{9AFF99} 
           & How to collect and integrate accessibility information in   existing geovisualizations?                            &     x                &                    &         x           \\ \hline
\rowcolor[HTML]{9AFF99} 
           & How to model user abilities and experience, and include these aspects into relevance rankings?                            &                     & x                   & x                   \\ \hline
\rowcolor[HTML]{9AFF99} 
           & How to automatically verify the accessibility (level) of web-based geovisualizations?                                        &                     &                     & x                   \\ \hline
\rowcolor[HTML]{9AFF99} 
           & How can accessibility information be gathered on-the-fly, and   integrated into relevance ratings?                 &                     & x                   & x                   \\ \hline
\rowcolor[HTML]{9AFF99} 
           & How to realize plastic geovisualizations?                                                                          &                     & x                   & x                   \\ \hline
\rowcolor[HTML]{9AFF99} 
           & How to encode accessibility information about web-based geovisualizations in a machine readable format? &   x                  &                     & x                   \\ \hline
\multicolumn{5}{|l|}{}                                                                                                                                                                            \\ \hline
\rowcolor[HTML]{CBCEFB} 
\textbf{I} &                                                                                                                    &                     &                     &                     \\ \hline
\rowcolor[HTML]{CBCEFB} 
           & To which extent are given symbols/palettes intuitive for users and to which extent do they differ across providers?                                    &                    &    x                 &                    \\ \hline
\rowcolor[HTML]{CBCEFB} 
           & How to formalize content of geovisualizations as   machine-readable statements?                                    & x                   &                     & x                   \\ \hline
\rowcolor[HTML]{CBCEFB} 
           & How to formalize user insights as machine-readable statements?                                                     & x                   &                    & x                   \\ \hline
\rowcolor[HTML]{CBCEFB} 
           & How to capture connotations of geovisualizations in a   machine-readable format?                                   & x                   &                     & x                   \\ \hline
\rowcolor[HTML]{CBCEFB} 
           & How to make map design conventions machine-accessible?                                                      &    x                 &                    &          x           \\ \hline
\rowcolor[HTML]{CBCEFB} 
           & How to model data gaps formally?                                                                                   & x                   &                     & x                   \\ \hline
\rowcolor[HTML]{CBCEFB} 
           & How to design interactive guidelines for geovisualization design?                                                &                     &             x        & x                   \\ \hline
\rowcolor[HTML]{CBCEFB} 
           & How to establish mutual understanding between web agents during geodata exchange tasks?             & x                   &                     & x                   \\ \hline
\multicolumn{5}{|l|}{}                                                                                                                                                                            \\ \hline
\rowcolor[HTML]{FFCCC9} 
\textbf{R} &                                                                                                                    &                     &                     &                     \\ \hline
\rowcolor[HTML]{FFCCC9} 
           & How to automatically infer a geovisualization licensing policy   from the licensing policies of its components?    & x                   &                     &        x             \\ \hline
\rowcolor[HTML]{FFCCC9} 
           & How to manage licensing policies of geovisualizations over   time?                                               &                    &                     & x                   \\ \hline
\rowcolor[HTML]{FFCCC9} 
           & How do geovisualization reuse approaches perform for different groups of users, and how to describe these approaches formally?                                        &                     & x                   &                     \\ \hline
\rowcolor[HTML]{FFCCC9} 
           & How to estimate transferability of geovisualization components  a priori?                                         &                     &                     & x                   \\ \hline
\rowcolor[HTML]{FFCCC9} 
           & What are developer needs and issues for reuse of   geovisualization components?                                    &                     &                     & x                   \\ \hline
\rowcolor[HTML]{FFCCC9} 
           & How to realize digital rights management and smart contracting for geovisualizations at a Web scale?              &                     &                     & x                   \\ \hline
\end{tabular}
\end{small}
\caption{Examples of open questions on the road towards FAIR geovisualizations. A cross indicates the perspectives for which the question are mostly relevant: C stands for computer and indicates knowledge representation issues; A stands for analyst and suggests user interface design issues; and D stands for developer and indicates issues related to technical considerations of publishing online geovisualizations.}
\label{tab:openquestions}
\end{table}

\subsection{Findable geovisualizations}
Realizing step 1 of the scenario needs work on `description', i.e. semantic annotation of existing online geovisualizations. Linked Data \citep{kuhn2014linkeddata} has been the de facto technique for semantic annotation of georeferenced data in recent years. Semantic annotation research needs mechanisms to produce large-scale annotations for existing geovisualizations on the Web. These annotations need to cope with the peculiarities of geovisualizations, i.e. annotations can be graphical, or textual or a mix of both (see \cite{Vanhulst2018}). Also, these annotations need to be fine-grained, that is, capture specifics of the data views (e.g. map, data table, bar chart) composing a geovisualization. Using form-based interaction to produce annotations could be daunting, and we could explore techniques that take advantage of recent progress in speech-based interaction and named entity recognition to enable users to accurately describe cartographic products on the Web. The descriptions should not only touch on missing metadata (e.g. author, creation date), but also inform about content-related aspects of the geovisualizations (e.g. what a user \textit{U} learned from a geovisualization \textit{G} during an interaction session \textit{I}). Work in this area could capitalize on the descriptions of the sub-components (e.g. datasets or software modules/functions/libraries used to produce the geovisualization), if available. Besides, these annotation mechanisms should not be disconnected from existing platforms (e.g. ESRI Storytelling, DataUSA, or the European Open Data Portal), but available `where the action is', that is, as easy-to-use add-ons within these platforms. Finally, techniques will be needed to maintain these crowdsourced annotations over time, and preserving the multiplicity of views while ruling out mischief.

`Persistence' is mostly relevant to step 2 of the scenario. A user does not want to click on a geovisualization that points to a ``HTTP 404 not found'' error. Several platforms (e.g. Zenodo, Spatial Data Hub, PURL\footnote{\url{www.purl.org}}) make it now possible to assign persistent URLs (Uniform Resource Locators) to documents on the Web. In the context of Linked Data, the Web community has proposed  PingtheSemanticWeb and the ``Web of Data – Link Maintenance Protocol" \citep{Volz2009} as preliminary ideas for protocols to manage the evolution of links on the Web. Traditional web-geovisualization development produces HTML/CSS/JavaScript code deposited on a server, and then accessed from a client that possesses a rendering engine (i.e. the browser). Evolving from this status-quo will necessitate an increasing adoption of a container-based approach to geovisualization development (coupled perhaps with continuous deployment). In addition, application programming interfaces (API) will play a key role, particularly when datasets are too big to be stored locally. For these cases, a persistent way of retrieving datasets at the API level would need to emerge. That is, a DOI-based approach to dataset retrieval is key to accessible geovisualizations. Platforms such as Dataverse\footnote{\url{https://dataverse.org/}} and DataCite\footnote{\url{https://datacite.org/}} allow the assignment of DOIs to datasets. DataCite enables also retrieval of these datasets according to their DOIs at the API level, and is an example implementation of the DOI-based approach to dataset retrieval. Thus, the challenges of persistent geovisualization ahead are more of social nature (i.e. how to ensure widespread adoption) than technical. Further collaborations between relevant institutions (e.g. OGC, the W3C) will be key in overcoming them. 

`Relevance' is also a notion that needs to be put under closer scrutiny. In particular, geographic information retrieval research needs to explore how the ranking of geovisualizations can be implemented to produce cognitively plausible results (e.g. how do factors affecting relevance ranking for geovisualizations differ from those used for ranking websites or datasets?). Finally, most graphical perception studies so far have only considered static geovisualizations, but interactivity should receive due attention on the road towards sensible rankings of findable geovisualizations. 

Example questions:
\begin{itemize}
    \item How to generate large-scale semantic annotations for web-based geovisualizations?
    \item How to enable fine-grained annotation of geovisualization components (e.g. data views, datasets, projection, mapping libraries)?
    \item How to record and share geographic insights across interaction sessions?
    \item How to model the relevance of online geovisualizations for the purpose of semantic search?
    \item How to realize effective aliasing for geovisualization search?
\end{itemize}


\subsection{Accessible geovisualizations}
As mentioned above, modeling user abilities/expertise, and realizing plastic geovisualization presents opportunities for further research on accessible geovisualization, i.e. tailoring to abilities and expertise, platform- and device-independence. There is also a need for tools translate accessibility information into machine readable format, wherever possible, so that these can be used during the implementation of geovisualizations. At the moment, there is still a disconnect between work on empirically derived recommendations regarding the accessibility of web maps (e.g. \cite{Hennig2017}) and work on technical means to document accessibility requirements for web-pages (e.g. \cite{Harper2007, Pelzetter2020}).

Example questions:
\begin{itemize}
    \item How to collect and integrate accessibility information in existing geovisualizations?
    \item How to model user abilities and experience, and include these aspects into relevance rankings?
    \item How to automatically verify the accessibility (level) of web-based geovisualizations?
    \item How can accessibility information be gathered on-the-fly, and integrated into relevance ratings?
    \item How to realize plastic geovisualizations?
    \item How to encode accessibility information about web-based geovisualizations in a machine readable format?
\end{itemize}




\subsection{Interoperable geovisualizations}
\label{subsubsec:interop}
Interoperability of geovisualizations (i.e. formalization of statements, compliance with widespread design conventions) is relevant to steps 2, 3 and 4 of the scenario. There is a need for user testing to evaluate actual user interpretation of map symbols, potential misunderstandings, and the reasons for these. This research will be mostly valuable for step 2. \textcolor{black}{Despite much research on the interpretation of map symbols (e.g. \cite{Korpi2010,Akella2009,Kinkeldey2014,Schnurer2020,Koylu2017}), formal rules regarding map symbolization are still not consolidated, perhaps due to the diversity of map context usages available and studied. Initiatives such as the GIS\&T Body of Knowledge (Section: Cartography \& Visualization, \url{https://gistbok.ucgis.org/}), which synthesize insights from the academic literature and communicate them in a language accessible to non-experts will thus play an important role for awareness-raising, the adoption of best practices, and the realization of more interoperable geovisualizations (analyst's perspective). In addition, advances on semantic interoperability research (e.g. formal specification of geodatasets' content and geovisualization components) is needed to realize steps 3 and 4 of the scenario.} So far, semantic interoperability research in GIScience has invested heavily in ontology building. These ontologies have been used to generate theories of geoinformation (e.g. \cite{couclelis2010ontologies, bittner2009spatio,agarwal2005ontological}), for query disambiguation and expansion (e.g. \cite{jones2004the,Lutz2006,Purves2007}), and for the (semi-automatic) composition of geoprocessing services (e.g. \cite{Lemmens2006}). Ontologies are also key to the Sensor Web Enablement initiative of the OGC (see \cite{broring2011new}). While these works offer valuable insight, \textit{demos of agents} that use these ontologies to automatically exchange things during a communication process have been less frequent. Ergo, GIScience research needs to go beyond ontology building to embrace \textit{semantic agent building}, where the agents select ontologies relevant to a given scenario, and exchange information with other relevant agents in an autonomous way, to solve some specific tasks. This is a major gap in current research. Identifying the tasks that cover scenarios of value in GIScience, and following up with agents that solve these tasks, will provide a much clearer measure of progress for semantic interoperability research than ontology provision alone. As to coping with missing data, formalizing and providing solutions to the issue has been relatively rare in geovisualization research. Notable exceptions are \textcite{Robinson2019}'s typology of strategies to draw users' attention to the presence of gaps in the data, and \textcite{ballatore2019context}'s context frame approach to generates visual cues about objects not shown on the map in order to minimize the need for zooming operations during an interaction session.

Moving to the point were semantic agents can seamlessly cooperate, needs advances along three axes: agents, protocols, and messages. The exact role of agents remains to be specified, but at least four roles can be envisioned: discovery (i.e. find out other visualizations on the Web that provide different perspectives on the data at hand); negotiation (e.g. establish with other visualizations on the Web whether or not data transfer would be possible); retrieval (i.e. get pieces of data from other visualizations); and integration (i.e. close data gaps of Geoviz1 with data items from Geoviz 2, see Figure \ref{fig:scenario}). Next to agent-oriented work, protocol-oriented work is needed to provide ways of exchanging messaging, building for example on state-of-the-art web technologies. Finally, message-oriented work can provide a uniform way of formally representing the content of datasets underlying the geovisualizations (e.g. their topic \& typical spatio-temporal attributes such as extent, temporal coverage, accuracy, and resolution), and of the geovisualizations themselves. \textcite{kuhn2012core}'s core concepts of spatial information may provide a starting point for a vocabulary for data content annotation; \textcite{roth2013empirically}'s taxonomy of interaction primitives for interactive geovisualization may also be useful for the annotation of geovisualizations. It remains to be seen whether these existing vocabularies/taxonomies are sufficient, or whether they should be extended (and in which way) to enable message exchanging for interoperable geovisualizations. Finally, a measure of more social nature on the roads towards interoperable geovisualizations could be the organization of (yearly) contests, ending up with demos showing geovisualizations exchanging some data autonomously (or deciding not to exchange data, based on a mutual understanding). \textcolor{black}{\textcite{Plaisant2008} reported positive experiences from annual competitions for the development of new visualization systems in the Information Visualization community}. 


Example questions:
\begin{itemize}
    \item To which extent are given symbols/palettes intuitive for users and to which extent do they differ across providers?
    \item How to formalize the content of geovisualizations as machine-readable statements?
    \item How to formalize user insights as machine-readable statements?
     \item How to capture connotations of geovisualizations in a machine-readable format?
    \item How to model data gaps formally?
    \item How to design interactive guidelines\footnote{\textcolor{black}{Interactive guidelines walk users through problem-solution patterns in an interactive manner, to transfer insights from the academic literature to non-experts \citep{trilles2020interactive}.}} for geovisualization design?
    \item How to establish mutual understanding between web agents during geodata exchange tasks?
\end{itemize}

\subsection{Reusable geovisualizations}
Reuse (i.e. licensing, transformability, and modularity) is primarily relevant to steps 5 and 6 of the scenario. Concepts such as `toolkits', `map use contexts', `intelligent geovisualization', and `map plasticity' have provided a good start to pinpointing the many dimensions of geovisualization reuse. Yet, we still need a clearer picture of reusability approaches available to geovisualizations creators and users, and most importantly, of the needs of users within each of the approaches. Put differently, a theoretical description of reusability approaches and a comparison of their merits is a challenging, yet much need contribution necessary to advance the current state-of-the-art on reusable geovisualization. As mentioned in Section \ref{subsec:fairuser}, template-based designs realize reuse across topics effectively: the same visualization template (e.g. choropleth, graduated symbol, cartogram) can be used across different topical domains (e.g. politics, demographics, real estate). The idea of template reuse in one-click (e.g. drag and drop) is appealing, but may be challenging to fully realize given the complexity of the design space of geovisualizations. Yet, it could be valuable during the comparison of geovisualization reuse approaches to assess research progress. For example, the number of clicks needed to create a visualization can be used along with established measures such as expressiveness (i.e. the number of geovisualization types that can be created), during the evaluation of tools and toolkits facilitating geovisualization reuse. \textcolor{black}{An early example of a drag-and-drop approach to visualize datasets was proposed in \citep{Wills2010}. The approach focused on detecting structural (as opposed to semantical) aspects of datasets, relies on the grammar of graphics, and was illustrated primarily on non-geographic datasets. Extending it to account at least for semantical and spatial aspects of datasets would be needed to fully realize steps 5 and 6 of the scenario.} Finally, a more social aspect of geovisualization reuse touches upon the licensing policy. By default, nothing is reusable unless permission is explicitly granted. This principle does not adequately reflect the state of matters on the Web. To give an example, the source code of web-based geovisualizations can be accessed through browser inspection tools. If the visualization has been developed with a library based on Web standards such as HTML, CSS (e.g. D3.js), data items powering the visualization can also be accessed. That is, licensing of web-based geovisualization may consider moving to a new motto: `everything is reusable unless reuse is explicitly restricted' to reflect openness of resources on the Web. 

Example questions:
\begin{itemize}
    \item How to automatically infer a geovisualization's licensing policy from the licensing policies of its components?
    \item How to manage licensing policies of geovisualizations over time?
 \item How do geovisualization reuse approaches perform for different groups of users, and how to describe these approaches formally?
    \item How to estimate the transferability of geovisualization components a priori?
     \item What are developer needs and issues for reuse of geovisualization components?
\end{itemize}

\subsection{Summary: what are open questions for FAIR geovisualizations research?}
\begin{figure}
    \centering
    \frame{
    \includegraphics[scale=0.6]{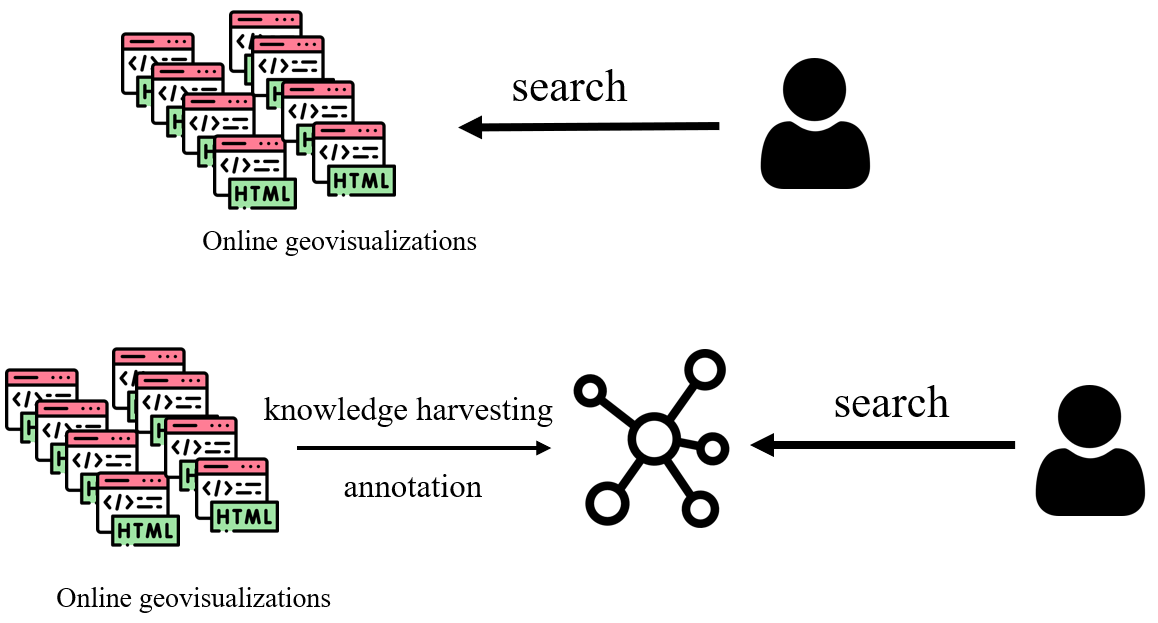}}
    \caption{An example of practical step towards more FAIR geovisualizations. Top: current online visualizations are treated as html documents for search; Bottom: exploring how knowledge harvesting techniques and semantic annotation can help build knowledge graphs for advanced search of geovisualizations at a Web scale.}
    \label{fig:practical}
\end{figure}

\textcolor{black}{Revisiting the scenario introduced at the beginning of the article, this section has contributed open questions on the road towards FAIR geovisualizations. The questions listed are by no means exhaustive, but illustrate that new types of problems arise to realize findability, access, interoperability, and reuse of web-based geovisualizations. Immediate, practical steps to make progress on steps 1-3 of the scenario could include the use of knowledge harvesting techniques \citep{Weikum2019} and of semantic annotation to generate knowledge graphs from geovisualizations stored as HTML-documents, the development of techniques to search for relevant geovisualizations over these knowledge graphs, and techniques to maintain online geovisualizations and their respective knowledge graphs synchronized (Figure \ref{fig:practical}).} \textcolor{black}{Using visualizations of the Observable gallery as a testbed could be helpful to start, as these are already annotated using the Open Graph Protocol\footnote{Thanks to one anonymous reviewer for pointing at Observable.}}.


\section{Limitations} 
\label{sec:limitations}
This article should be thought of as an initial exploration of FAIR geovisualizations, not the final word. There are some limitations that can be noted. 

First, the work has argued that geovisualizations should be considered as \textit{a unit in their own right}, for geographic information search and reuse on the Web. Getting there necessitates more work on making geovisualizations \textit{fair} (in the literal sense) to the datasets they depict. There are indeed numerous factors that affect visual encodings that may impede the fairness of visualization with respect to the data they portray. These include, among others, graphical inference (are the patterns we see really there?, see \cite{wickham2010graphical}), Tufte lie factor (is the size of the effect shown in the graphic directly proportional to the size of the effect in the data), perceptual effectiveness (is the visual symbol to encode differences appropriately chosen?), and imperfections of data classification \citep{Monmonier2005}. These issues, though acknowledged, are not discussed in detail here. Accounting for errors occurring in data-to-visuals transformation processes, \textcolor{black}{and more broadly of uncertainties inherent to data exploration processes and their visual communication}, can be examined once prototypical applications emerge that realize the steps presented in Figure \ref{fig:scenario}. 

Second, as to the formulation of the open questions, GIScience has often used forecasting to articulate research agendas. For example, forecasting has been used to predict the evolution of spatial data infrastructures in \citep{diazfuture}, anticipate the development of the digital earth in \citep{Goodchild2012a, Craglia2012}, and conjecture developments of the field as a whole in \citep{Goodchild2009,goodchild2010twenty}. We've learned that forecasting may miss important developments (for an anecdote, see \cite{Goodchild2009}), and backcasting too may miss important developments of the years to come. Finally, similar to forecasting, backcasting is `value-driven' \citep{Dreborg1996}, that is, the ideas are inevitably biased towards the author's research.


\section{Conclusion}
\label{sec:conclusion}
This work has applied the FAIR principles to a new domain (i.e. geovisualization), and taken a holistic approach to FAIRness to uncover some challenges and opportunities for GIScience research. This article contributes to geovisualization research in several ways. First, the work highlighted that geovisualizations are information products increasingly available, but sufficiently distinct from websites and raw datasets to deserve sustained effort aiming at making them FAIR (Findable, Accessible, Interoperable and Reusable). Second, the work discussed three complementary perspectives on FAIR geovisualizations (the computer, the analyst, and the developer), and clarified their respective demands. The framework resulting from the discussion can help researchers involved in realizing FAIR geovisualization to disambiguate their stance. It can also be used to systematically map research on FAIR geovisualizations as it evolves. Third, relevant approaches from the literature to realize FAIRness and some unique challenges of FAIR geovisualizations were brought forth and discussed. At last, the article sketched open questions towards FAIR geovisualizations. These questions arguably present huge challenges for current research but challenges worth attending to. The future of geovisualization research and practice is FAIR.


\section*{Acknowledgements}
I thank several anonymous reviewers for their detailed and very insightful comments on earlier versions of this article. The icons in Figures \ref{fig:scenario} and \ref{fig:practical} were taken from \url{www.flaticon.com} (designers: `Freepik', `Kiranshastry', `Nhor Phai', `Pixel perfect', and `srip').

\section*{Supplementary material}
\label{sec:supplement}
The supplementary material is available at \url{https://doi.org/10.6084/m9.figshare.13635563}.


\printbibliography
\end{document}